\documentclass[conference,10pt]{IEEEtran}
\usepackage[left=1.66cm,right=1.66cm,top=1.8cm,bottom=4.21cm]{geometry}
\IEEEoverridecommandlockouts
\usepackage{cite}
\usepackage{amsmath,amssymb,amsfonts,bbm}
\usepackage{amsthm}
\usepackage{graphicx}
\usepackage{textcomp}
\usepackage[dvipsnames]{xcolor}
\usepackage{multirow}
\usepackage[nolist]{acronym}

\usepackage{siunitx}
\usepackage[acronym,nomain,nonumberlist]{glossaries}
\usepackage{subcaption}
\usepackage{caption}
\usepackage{tikz}
\usetikzlibrary{matrix,shapes}
\usepackage{tkz-tab}
\usetikzlibrary{automata,arrows,positioning,calc}
\usetikzlibrary{shapes,snakes}
\usetikzlibrary{fit}
\usepackage{dsfont}
\usepackage{soul}
\usepackage{subfiles}
\usepackage{comment}

\usepackage{algorithm}
\usepackage{algpseudocode}
\makeatletter 
\algrenewcommand\ALG@beginalgorithmic{\footnotesize}
\makeatother

\usepackage{booktabs}

\newcommand{\figref}[1]{\figurename~\ref{#1}}

\colorlet{pink}{red!40}
\colorlet{blue}{cyan!50}

\tikzset{
  treenode/.style = {shape=rectangle, rounded corners,
                     draw, anchor=center,
                     align=center,
                     top color=white, bottom color=blue!20,
                     inner sep=1ex},
  decision/.style = {treenode, rectangle,  font=\normalsize, inner sep=1ex},
  root/.style     = {treenode, font=\normalsize, bottom color=blue!20,inner sep=1ex},
  env/.style      = {treenode, font=\ttfamily\normalsize},
  finish/.style   = {root, bottom color=green!40},
  dummy/.style    = {circle,draw}
}

\usepackage{float} %to prevent tables repositioning

\colorlet{pink}{red!40}
\colorlet{blue}{cyan!50}
\begin{document}

% This code is to reduce the list of authors by using et. al:
\bstctlcite{IEEEexample:BSTcontrol}

\title{Cluster-Specific Predictive Modeling: A Scalable Solution for Resource-Constrained Wi-Fi Controllers}
% \title{An Energy-oriented Time Series Clustering and Prediction Approach in Constrained Centralized Wi-Fi Networks}
%\title{Enhancing Wi-Fi Network Performance and Efficiency via Feature-Based Clustering and Specialized Predictive Models}

\author{
\IEEEauthorblockN{Gianluca Fontanesi$^{\star}$, Luca Barbieri$^{\star}$, Lorenzo Galati Giordano$^{\star}$, Alfonso Fernandez Duran$^{\flat}$ and Thorsten Wild$^{\star}$ \vspace{0.1cm}
}
\IEEEauthorblockA{$^{\star}$\emph{Radio Systems Research, Nokia Bell Labs, Stuttgart, Germany}}
\IEEEauthorblockA{$^{\flat}$\emph{Nokia Spain, Madrid}}
\IEEEauthorblockN{\thanks{This work is partially supported by UNITY-6G project, co-funded from European Union’s Horizon Europe Smart Networks and Services Joint Undertaking (SNS JU) research and innovation programme under the Grant Agreement No 101192650 and from the Swiss State Secretariat for Education, Research and Innovation (SERI). This work is also partially supported by CDTI project IDI-20250211 MINERGY. Corresponding author: \emph{gianluca.fontanesi@nokia-bell-labs.com}.}}
}

% Contact information of all authors: \emph{luca.barbieri@nokia-bell-labs.com} \emph{lorenzo.galati$\_$giordano@nokia-bell-labs.com}, \emph{thorsten.wild@nokia-bell-labs.com}.

\maketitle
%%%%%%%%%%
% FORCE PAGE NUMBERS
% \thispagestyle{empty} %plain
% \pagestyle{plain}
%%%%%%%

\begin{abstract}
This manuscript presents a comprehensive analysis of predictive modeling optimization in managed Wi-Fi networks through the integration of clustering algorithms and model evaluation techniques. The study addresses the challenges of deploying forecasting algorithms in large-scale environments managed by a central controller constrained by memory and computational resources. Feature-based clustering, supported by Principal Component Analysis (PCA) and advanced feature engineering, is employed to group time series data based on shared characteristics, enabling the development of cluster-specific predictive models. Comparative evaluations between global models (GMs) and cluster-specific models demonstrate that cluster-specific models consistently achieve superior accuracy in terms of Mean Absolute Error (MAE) values in high-activity clusters. The trade-offs between model complexity (and accuracy) and resource utilization are analyzed, highlighting the scalability of tailored modeling approaches. The findings advocate for adaptive network management strategies that optimize resource allocation through selective model deployment, enhance predictive accuracy, and ensure scalable operations in large-scale, centrally managed Wi-Fi environments.
\end{abstract}

% Continuous Channel Access, Unlicensed Spectrum

\newacronym{ap}{AP}{Access Point}
\newacronym{aiml}{AI/ML}{Artificial Intelligence and Machine Learning}
%\newacronym{ai}{AI}{Artificial Intelligence}
%\newacronym{ml}{ML}{Machine Learning}
\newacronym{svm}{SVM}{Support Vector Machine}
\newacronym{arima}{ARIMA}{Autoregressive Integrated Moving Average}
\newacronym{ann}{ANN}{Artificial Neural Network}
\newacronym{rnn}{RNN}{Recurrent Neural Network}
\newacronym{lte}{LTE}{Long-Term Evolution}
\newacronym{mae}{MAE}{Mean Absolute Error}
\newacronym{mape}{MAPE}{Mean Absolute Percentage Error}
\newacronym{rmse}{RMSE}{Root Mean Squared Error}
\newacronym{mlp}{MLP}{Multi-Layer Perceptron}
\newacronym{sta}{STA}{Station}
\newacronym{lstm}{LSTM}{Long Short-Term Memory}
\newacronym{cnn}{CNN}{Convolutional Neural Network}
\newacronym{svr}{SVR}{Support Vector Regression}
\newacronym{gnn}{GNN}{Graph Neural Network}
\newacronym{mac}{MAC}{Medium Access Control}
\newacronym{cdf}{CDF}{Cumulative Distribution Function}
\newacronym{pca}{PCA}{Principal Component Analysis}
\newacronym{wcss}{WCSS}{within cluster sum of squares}
\newacronym{kpi}{KPI}{key performance indicator}
\newacronym{csm}{CSM}{cluster specific model}

\section{Introduction}

As data from real-world operational networks becomes increasingly accessible, there is a growing need for research focused on effectively leveraging this data for network optimization and management. Wi-Fi solution providers are particularly interested in advanced analytical tools to enhance their products, yet progress in this area has been limited. Addressing this gap, our work investigates the use of clustering and forecasting algorithms to unlock the potential of operational network data while addressing critical resource constraints inherent to large-scale environments that are managed by a central controller with limited memory and computational resources.

Forecasting algorithms play a vital role in improving energy efficiency, performance, and reliability in Wi-Fi networks~\cite{chen2021flag}. Modern WLAN solutions often rely on centralized controllers that collect time series data from multiple \glspl{ap}. This data, periodically updated, can be utilized by forecasting algorithms to enable proactive decision-making. By predicting future trends and behaviors, these algorithms empower Wi-Fi networks to dynamically adapt to changing conditions~\cite{wilhelmi2024ai}. However, the application of forecasting algorithms in large-scale environments faces significant challenges due to the resources intensity needed and the associated costs.

Traditional univariate forecasting techniques, such as the \gls{arima} model, are commonly used for wireless prediction due to their ability to handle individual time series with limited data~\cite{ho1998useARIMA}. While effective for small-scale applications, \gls{arima} struggles to model multiple time series efficiently~\cite{siami2018comparisonLSTM_ARIMA}. To address this, global forecasting models, such as \gls{cnn} and \gls{rnn}, have been introduced to leverage information across sets of related time series \cite{wilhelmi2024ai}. These models, while powerful, often require substantial computational resources and memory, making them less suitable for deployment in industrial systems with stringent constraints.
 
% technical problem and state of the art

% Time series clustering can be broadly categorized into three main approaches: distance-based, feature-based, and model-based methods. Distance-based clustering algorithms operate directly on raw data points, using predefined distance metrics to measure similarity or dissimilarity between time series shapes~\cite{paparrizos2015k}. Feature-based clustering, on the other hand, relies on extracting descriptive features from the raw data to summarize its key characteristics, while model-based clustering uses mathematical models built from the raw data to group time series based on their underlying dynamics~\cite{liao2005clustering}. The choice of clustering approach depends heavily on the intended purpose and the nature of the dataset.
Time series clustering can be broadly categorized into three main approaches: distance-based~\cite{paparrizos2015k}, feature-based, and model-based methods~\cite{liao2005clustering}. Feature-based clustering extracts descriptive features from raw data to summarize key characteristics. Unlike distance-based methods, feature-based clustering is resilient to missing or noisy data and focuses on capturing overarching patterns rather than individual data points \cite{hyndman2015large}. This robustness and interpretability make feature-based clustering an ideal choice for cloud-based network controllers operating under storage and computational constraints.

% If the goal is to discriminate between the geometric profiles of time series, a shape-based algorithm such as dynamic time warping is suitable. In contrast, feature-based dissimilarity is preferable for comparing underlying quality characteristics.

% The notion of dissimilarity (or similarity) plays a central role in clustering algorithms, as it determines how time series are grouped. For instance, if the goal is to differentiate time series based on their geometric profiles, shape-based algorithms such as dynamic time warping are well-suited. These algorithms focus on aligning time series to minimize differences in their shapes. However, when the objective is to compare underlying quality characteristics, feature-based clustering is preferable, as it provides a higher level of abstraction and interpretability.

% Distance-based clustering approaches are highly sensitive to the choice of distance metric, which must account for factors such as noise, series length, dynamics, and scaling. Defining an appropriate distance measure can be challenging, as many metrics focus primarily on the shape of the time series and may produce unintuitive results, particularly when dealing with noisy or incomplete data. Additionally, the performance of distance-based methods often depends on the selected time window, which can significantly impact the clustering outcome \cite{xing2021wireless}. These limitations make distance-based clustering less suitable for datasets with complex or heterogeneous time series.
In this paper, we consider a Wi-Fi cloud-based network controller that manages a large number of \glspl{ap} under storage and computational constraints.
As the number of \glspl{ap} increases, the scalability challenge becomes more pronounced, with the potential number of clusters and predictive models growing significantly. 
Therefore, it is desirable to strike a balance between the deployed models (in terms of numbers, complexity and training) and resource utilization (e.g. memory) in the cloud.
% The process involves two main stages: feature extraction and clustering. During the feature extraction phase, features such as trends, autocorrelation, and seasonality are derived to represent the time series. In the clustering phase, standard algorithms like k-means are applied to group the time series based on these features.

Recent research has explored various ways to combine clustering and deployed models. The approach proposed in \cite{bandara2020forecasting} is to first divide the time series using clustering techniques (e.g. feature-based), and then fit available global models considering the series within each cluster. Alternatively, the work in \cite{lopez2025time} proposes grouping the series based on the minimum resulting forecast error of the assigned global model. The clustering algorithm is specifically designed to allocate the different time series so that the corresponding models best represent the existing prediction patterns.
Both deploying a separate predictive model for each cluster or using highly complex models across all clusters may be infeasible for constrained storage and computational resources centralized controllers.

%proposed method
% a method that iteratively establishes a dynamic mapping between clusters and predictive models. The approach begins with a step to exploit similarities between time series by discovering clusters of similar series. 
Exploiting clustering techniques, we propose an energy-efficiency-oriented deployment for constrained Wi-Fi network controllers. 
By identifying clusters with similar behaviors, the algorithm minimizes the need for maintaining multiple redundant models, thereby optimizing memory usage. 
Additionally, the clustering framework ensures that computational resources are allocated efficiently, avoiding the unnecessary execution of high-complexity models in low-activity or less critical clusters.

\section{System model}
The proposed framework assumes a Wi-Fi network consisting of \textit{I} \glspl{ap}. Measurements are collected from the different \textit{I} \glspl{ap} with a periodicity \textit{T} over a total of \textit{M} months or \textit{D} days.
We consider a network where the network measurements collected at the \glspl{ap} are sent to a central entity and stored in a centralized database located in a cloud data management system that contains the historical measurements for the different \glspl{ap}.
On this data, an offline clustering and traffic prediction training can be performed in order to extract knowledge from the collected data. It can be assumed that these offline procedures can be performed with certain periodicity in order to keep updated the obtained model according to recent collected measurements. 
The prediction phase can be performed online for the whole network or by deploying different models for the different clusters. During the prediction phase, new measurements can be utilized to confirm the accuracy of the training.

We consider a set of $I$ time series, $\mathcal{S}= \{X_{1},X_{2},...,X_{I}\}$, where the time sequence of a specific collected metric of the \textit{i}-th \gls{ap} is represented as $X_i = \{ x_{i,1}, x_{i,2}, ...,x_{i,N}\}$ where $\textit{i}=1,2,...,I$ and $N$ is the total number of samples for each \gls{ap} collected during the network measurement window.
% each AP should have more than one time series $\mathcal{S}= \{X_{i}^{1},...,X_{i}^{n}\}$ but for simplicity...
% (without loss of generality, it can be assumed that all series have the same length).
% We assume that each series $X_i$ contains training and validation sets.
 % where each $\mathcal{X}_{t}^{i} = (X_i^{(i)},...,X_{L_i}^{(i)})$, $i=1,...,n$.
First, we perform clustering on the elements of $\mathcal{S}$ and, second, construct and train a GM on the entire set of time series. 
% We will provide more details on the clustering technique in the next section.
In what follows, we clarify the meaning and the creation of the GM, that serves as a baseline for traffic prediction and is designed to provide generalized predictions across the network.
% The trained GM is used as default option for inference on the \gls{ap}. The global model determine the number of models needed to enhance the performance of traffic load prediction for each cluster that does not meet the required performance % We propose to consider a global model as which is fitted to the series in $\mathcal{S}$ We consider a global model as a learning algorithm able to fit the same forecasting function to all the time series in a set. 

\subsection{Global and Cluster Specific Models}
Global models are learning algorithms that fit the same forecasting function to all time series in a set.
They are in contrast with local models, which adjust a different function to each time series in the database \cite{lopez2025time}.
Recalling $\mathcal{S}$ as the collection of all sets of time series, $\mathcal{V}^P = \{v_1,v_2,...,v_p\}$ is the vector of finite length representing the observed time series. We are interested in the future part of each series up to $h$ time steps, which can be seen as a vector of $\mathcal{V}^h = \{v_1,v_2,...,v_h \}$.
To compute the corresponding predictions, we employ a forecasting function $f$, which maps the observed part of the time series $\mathcal{V}^P$ to the future part $\mathcal{V}^h$, i.e., $f: \mathcal{V}^P \rightarrow \mathcal{V}^h$.
% which is a function from the observed time series to the future part, i.e. $f: \mathcal{V}^P \rightarrow \mathcal{V}^h$. 
A global model (or a set of), GM, is a learning algorithm that uses all time series in $\mathcal{S}$ to produce a single forecasting function $GM$.The global model directly maps the observed part of the time series $\mathcal{V}^P$ (from all series in $\mathcal{S}$) to the future part $\mathcal{V}^h$, i.e., 
$
GM: \mathcal{V}^P \rightarrow \mathcal{V}^h.
$
% Based on our previous results in \cite{wilhelmi2024ai}, we consider that the global model GM is constructed using a classical regression model. 
Due to our previous results in \cite{wilhelmi2024ai} on the same dataset, we select \gls{lstm} model as reference GM model.
\gls{lstm} is a type of \gls{rnn} designed to address long-range dependencies in sequential data thanks to the usage of \gls{lstm} cells, which allow capturing temporal dependencies across time series data. In this work, the time series of network data are passed through a few \gls{lstm} layers before a predictive output is generated by a feed-forward fully-connected layer.

In addition, we consider a number of clusters $K$ as a result of a clustering function $\mathcal{C}$ such that $S \rightarrow \{C_1,...,C_K \}$.
Assume there are $n_k$ series in the $k$th group $C_k$, i.e. $C_k= \{X_{k_{n1}},...,X_{k_{n_k}}\}$, obtained from a clustering process, where $k=1,...,K$ and the subscript $k$ is used to indicate that the corresponding series belong to cluster $k$.
% It is expected that the predictive ability of model GM with respect to the series in cluster $C_k$ better the more related the series in the group are.
A \gls{csm} learning algorithm (or a set of) $L$, is a function that takes a time series $C_k$ and returns a different forecasting function for each $C_k$ such that $L: \mathcal{V}^P \rightarrow \mathcal{V}^h$.
% trained independently for each cluster Ck

\section{Proposed clustering-based prediction framework}\label{Procedure}
\begin{figure}[t!]
\centering
\includegraphics[width=\linewidth]{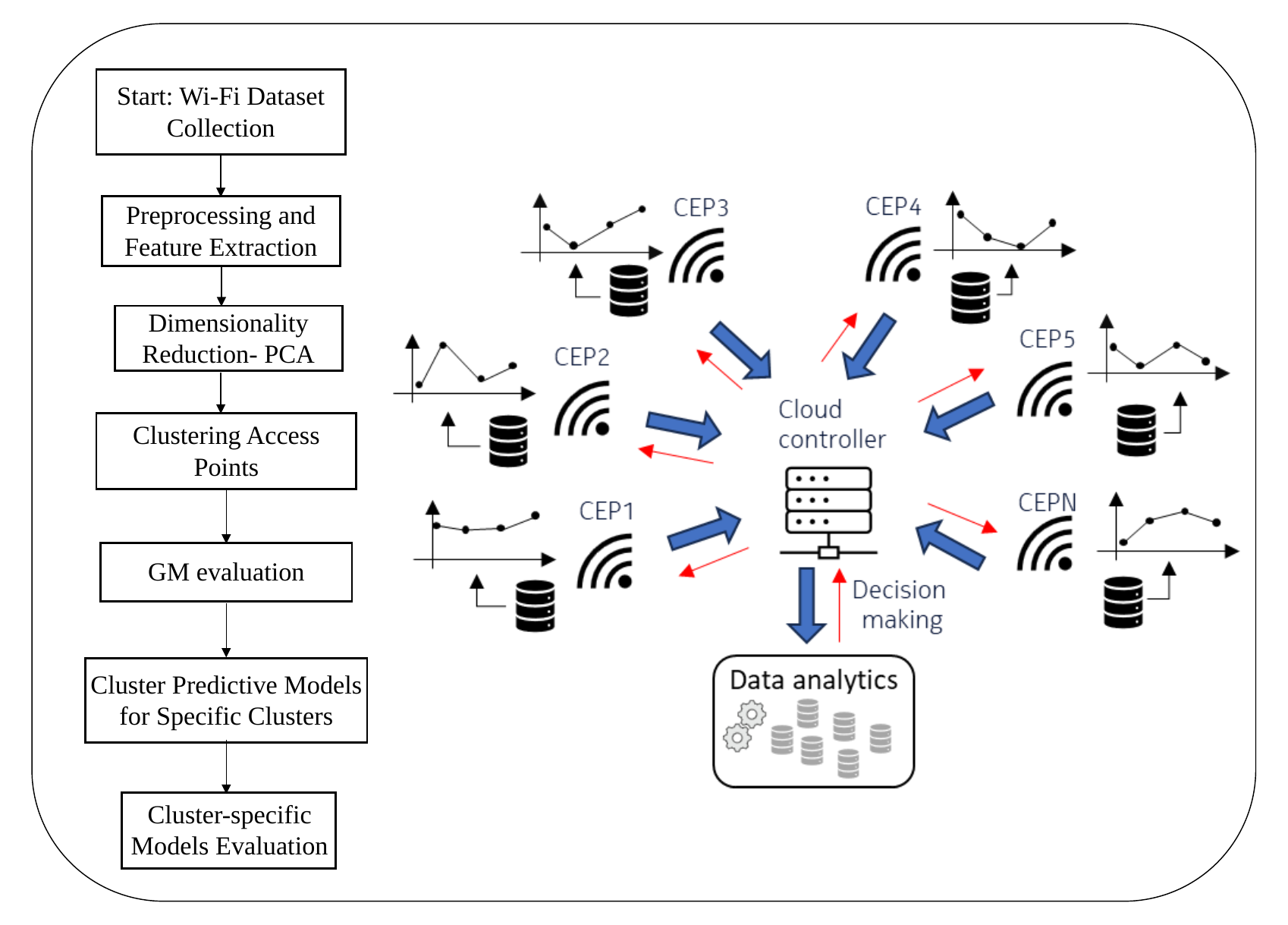}
\vspace{0.1cm}
    \caption{Proposed clustering and prediction analysis framework.}
    \label{fig:Procedure}
\end{figure}

Resource constraints such as limited memory and computational power pose significant challenges for deploying predictive models at scale. To address these challenges, we propose a clustering-based framework that enables efficient model training and selection in the controller by grouping similar time series data into clusters. This approach optimizes resource allocation and reduces computational overhead, making it particularly suitable for centralized controllers operating under stringent constraints.

To validate this approach, we aim to compare the performance of a GM trained on the entire dataset against the performance of the same \gls{lstm} model, but trained only on the \glspl{ap} belonging to specific clusters. Clustering serves as an augmentation step to exploit the similarity between time series, enabling tailored model training for each cluster.

% We propose a feature-based clustering approach using a set of interpretable features of a time series to obtain meaningful clusters.
% Note that the number of clusters ($K$) can be selected running the clustering algorithm

\subsection{Time series Clustering}\label{timeSeriesClustering}
The clustering process consists of feature extraction and clustering. A set of features $f$ such as trends, autocorrelation, and seasonality are derived to represent the time series $\mathcal{S}$, emphasizing energy-related dynamics like low activity patterns. 
% These features capture the dynamics of the data while emphasizing energy-related characteristics, such as patterns of low activity during specific periods. 
Two feature extraction strategies exist: generating a large set of features to capture all possible dynamics or selecting a limited set of interpretable features tailored to the application. Our framework prioritizes the latter, focusing on energy efficiency by selecting self-describable features that capture observable dynamics (see Table \ref{tab:features_grouped}).
% \Comment{[TW: As you mention energy efficiency now, maybe it's good if you spent a sentence in the introduction on it for describing it as one of the goals for usage of the prediction.]}

Preprocessing steps include \textbf{data transformation}: techniques such as logarithmic or cube root scaling are used to improve normality and homogeneity of variance, enabling algorithms to handle large variations in data effectively.
% Based on the connection timestamp, we create a new feature that classify each entry of the dataset in Weekend or Week, and day/afternoon/night.
The 'Byte' feature is further divided into quantiles to create categorical features ('low bytes', 'medium bytes', 'high bytes').
Data collected in university campuses exhibit distinct patterns on weekdays compared to weekends \cite{sone2020_TrafficPrediction_OJ-COMS}. New features are introduced (\textbf{feature engineering}), such as time-of-day classifications (e.g., 'Morning', 'Afternoon', 'Evening') and distinctions between weekdays and weekends, which capture temporal patterns relevant to energy efficiency and compute the average and standard deviation for each identified timezone.  We extract 35 features across three categories (Table~\ref{tab:features_grouped}): 
(i) \textit{Global statistics}: bytes and active users per AP (mean, std, quantiles); 
(ii) \textit{Temporal features}: bytes and user counts stratified by period 
(morning/afternoon/night) and day type (weekday/weekend), capturing both mean 
and variability ($2 \times 2 \times 3 \times 2 = 24$ features); 
(iii) \textit{Usage patterns}: peak-hour ratios and off-peak indicators.
% Additionally, we analyze the distribution of the 'Byte' feature to assess skewness and consider applying data transformations, such as logarithmic or cube root scaling, to address it.
Finally, features are scaled to ensure fair comparisons across different numerical ranges, preventing large-scale features from dominating smaller ones. 
In the clustering phase, standard algorithms like k-means are applied to group the time series based on these features.
% This helps effectiveness of k-means clustering depends on the distance matrix used to group data points, which requires features to be scaled appropriately. 
% To prevent features with large numerical ranges from dominating those with smaller ranges, normalization is applied to the dataset. 
% This ensures that all features have a mean of 0 and a standard deviation of 1, enabling fair comparisons across different scales.
The resulting features (see Table \ref{tab:features_grouped}) are designed to capture the majority of observable dynamics in the time series, including operational characteristics such as low-activity patterns during specific periods, which help identify 
clusters suitable for resource-efficient modeling.
To reduce the high dimensionality produced by this feature extraction process, \gls{pca} \cite{ding2004_ICML_PCA} is applied. This dimensionality reduction technique addresses feature redundancy by identifying the most informative relationships within the data, enabling meaningful clustering while reducing computational demands.

This framework achieves scalability by enabling selective model deployment: 
lightweight models serve simpler clusters, while specialized predictive models are deployed only for high-activity clusters where their complexity is justified. 
By leveraging clustering, the framework avoids the  computational burden of training and deploying complex models across all time series, making it ideal for large-scale, resource-constrained environments.

\begin{table}[t]
\centering
\caption{Extracted Features for Time Series Clustering (35 total)}
\small
\begin{tabular}{lc}
\toprule
\textbf{Feature Category} & \textbf{Count} \\
\midrule
\textbf{Bytes Statistics} & 3 \\
\quad Mean, Std, Quantiles (per AP) & \\
\midrule
\textbf{Active Users Statistics} & 2 \\
\quad Mean, Std (per AP) & \\
\midrule
\textbf{Temporal Features} & 30 \\
\quad \textit{Bytes:} Mean, Std $\times$ 3 periods\textsuperscript{†} $\times$ 2 day types\textsuperscript{‡} & 12 \\
\quad \textit{Active Users:} Mean, Std $\times$ 3 periods $\times$ 2 day types & 12 \\
\quad \textit{Usage Patterns:} Peak hours, off-peak ratios & 6 \\
\bottomrule
\multicolumn{2}{l}{\textsuperscript{†}Morning, Afternoon, Night; \textsuperscript{‡}Weekday, Weekend}
\end{tabular}
\label{tab:features_grouped}
\end{table}
% \gls{pca} partitions the variation in the data and identifies the most informative relationships, which can be used to create new features or reduce the dimensionality of the dataset. By focusing on the principal components, \gls{pca} enhances the clustering process by highlighting the most relevant patterns and relationships within the data, ultimately improving the interpretability and accuracy of the clustering results.
% PCA is a tool for dimension reduction in high dimensional data. A principal component is a combination of the original variables after a linear transformation. For example the first principal component captures the maximum variation in the rows of the m × n matrix. Therefore, loosely speaking the first k principal components capture the k most prevalent patterns in the data.

\subsection{Predictive framework}
% The \gls{lstm} GM is run for the first time to obtain the inference results for the \glspl{ap} belonging to different clusters ($K$). Additionally, the average performance across the entire dataset is obtained.
% The GM's performance is then assessed for each cluster and compared to its performance on the entire dataset.
% Next, the GM's performance is compared with that of the same \gls{lstm} model, trained only on the \glspl{ap} belonging to specific clusters.
% The \gls{lstm} model is trained on the \glspl{ap} that form the identified cluster ($k$), resulting in a new cluster specific trained model ($L_k$).
% Once trained, the new model ($L_k$) is used to perform inference on the selected cluster ($C_x$).
% The Performance Table is updated with the results of the new inference.
% In this work we don't select a specific threshold value but we generally compare the performance of the two approaches. 
To evaluate the effectiveness of the proposed clustering-based framework, a comparison is conducted between the \gls{lstm} (GM) and cluster-specific \gls{lstm} models ($L_k$). The process begins with the execution of the \gls{lstm} Global Model, GM, on the entire dataset to obtain inference results for all \glspl{ap} across different clusters (K). The GM's performance is assessed for each individual cluster.
% allowing for a comparison of the GM's performance on each cluster against its average performance across the entire dataset. 
This step highlights variations in predictive accuracy across clusters.

Once the GM's performance has been evaluated, cluster-specific models are trained for each identified cluster. These models are created by training a new \gls{lstm} model using only the \glspl{ap} belonging to the specific cluster $k$, that we will refer as $L_k$ in the rest of the paper. The training process results in cluster-specific models tailored to the unique characteristics of each cluster. After training, the cluster-specific model $L_k$ is used to perform inference on the corresponding cluster, and the inference results are recorded for further analysis.

The results obtained from the cluster-specific models $L_k$ and the related models memory occupancy are recorded in a Performance Table, which is stored in the controller, ensuring a comprehensive view of the models performance across all clusters. Finally, the performance of the GM and the cluster-specific models $L_k$ is compared. This comparison is conducted without applying a specific threshold value, focusing instead on a general evaluation of the two approaches. Metrics such as Mean Absolute Error (MAE), and 99th Percentile Absolute Error are analyzed to identify clusters where the GM performs poorly and assess the relative benefits of cluster-specific models.
The average prediction error with respect to the test set is measured 
    \begin{equation}\label{eq:errorPrediction}
        \frac{1}{n}\sum^K_{k=1}\sum_{i=1}^n d^*(\mathbf{X}_t^{(i)*},\hat{\mathcal{M}_k})
    \end{equation}
where $d^*$(.,.) is any function measuring discrepancy between the true values of $\mathbf{X}_t^{(i)*}$ and their predictions according to a model $\hat{\mathcal{M}_k}$ and $k$ the $k$-th cluster. If \gls{mae} is chosen as the error metric and $\hat{\mathcal{M}_k}$ is GM, then \eqref{eq:errorPrediction} becomes
\begin{equation}\label{eq:errorPrediction}
        \frac{1}{n}\sum^K_{k=1}\sum_{i=1}^n d_{MAE}^*(\mathbf{X}_t^{(i)*},\hat{\mathcal{M}_k})
    \end{equation}
with 
    \begin{equation}\label{eq:MAEerrorPrediction}
        d_{MAE}^*(\mathbf{X}_t^{(i)*},\hat{\mathcal{M}_k}) = \frac{1}{h} \sum_{j=1}^h |X_j^{i*}-\hat{F}^{i*}_{j,k}|,
    \end{equation}
where $\hat{F}^{i*}_{j,k}$ is the prediction of $X_j^{i*}$ according to the GM $\hat{\mathcal{M}_k}$.
% \footnote{It is worth highlighting that, if all the time series in the set are recorded in the same scale, then employing the MAE leads to meaningful conclusions. However, if that is not the case, \eqref{eq:MAEerrorPrediction} is likely to be a misleading performance measure, since series taking higher values are expected to have a larger impact on the computation of the average prediction error~\cite{lopez2025time}. This issue can be avoided by considering alternative error metrics}.
This methodology provides a \textbf{scalable} framework for evaluating the impact of clustering on prediction accuracy and memory usage, enabling the controller to adaptively allocate modeling resources: simpler clusters are served by the GM, while cluster-specific models $L_k,L_kv2$—though potentially heavier—are 
deployed only for challenging clusters where specialized modeling is justified by significant accuracy gains. This selective deployment strategy ensures the system scales efficiently to large scale network environments.

\section{Results}
In this section, we evaluate the proposed framework on a open source dataset~\cite{chen2021flag}. 
% First, based on the connection timestamp, we create a new feature that classify each entry of the dataset entry presented in Section \ref{datasetOpenSource} in Weekend or Week, and day/afternoon/night.
% We thus introduce a quantile byte feature that divide the bytes in three type of categorical feature like low Bytes, medium Bytes and high bytes. 
\subsection{Dataset}
We use the open-source dataset published in~\cite{chen2021flag}, which contains 25,074,733 association records from a total of 55,809 users extracted from 7,404 Wi-Fi \glspl{ap} distributed over a 3 km² campus area over 49 days in 2019. The data contains features such as the user's Media Access Control (MAC) address, the ID of the \gls{ap} to which the user is connected, and the number of bytes transferred per session (see \cite{chen2021flag} for more details).

Taking the association records from the original dataset, we derived the temporal uplink and downlink load of each AP by aggregating the contributions of each client individually. In particular, we equally spread the total load $\rho^i$ of each connection $j$ by the number of time steps of size $w$ in the duration of the association $T^j$ where $w$ is a configurable parameter that defines the granularity of the measurements (e.g. 1 s, 1 min, 1 h).

\subsection{Clusters visualization}
\begin{figure}[t!]
\centering
\includegraphics[width=\linewidth]{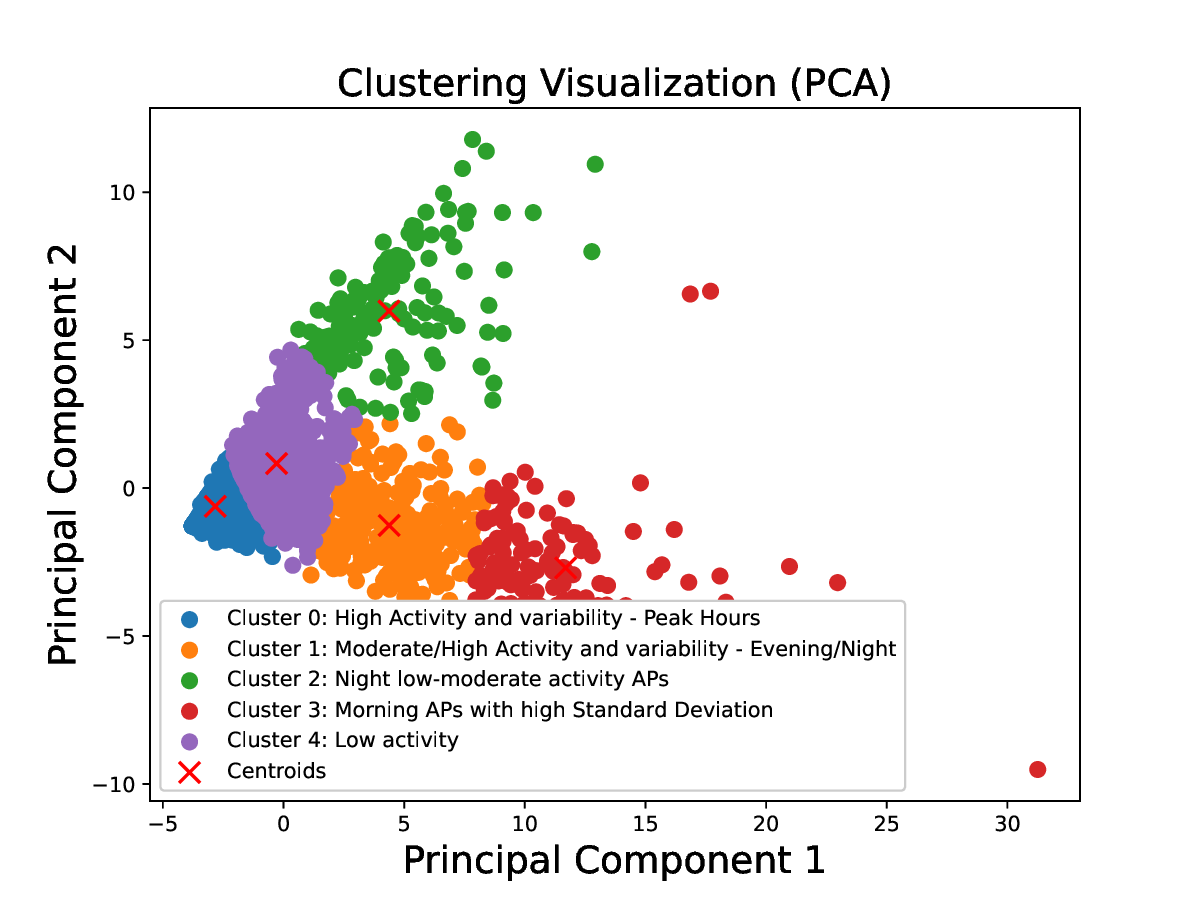}
\vspace{0.1cm}
    \caption{Clusters visualization.}
    \label{fig:cluster_vs_pca_plot}
\end{figure}

The application of Principal Component Analysis (PCA) followed by feature-based clustering, as described in Section \ref{timeSeriesClustering}, resulted in the identification of five distinct clusters using the k-means algorithm. The clustering validity was confirmed through established metrics such as the Calinski-Harabasz index and silhouette score, which demonstrated the robustness of the clustering approach. The first two principal components, along with the corresponding five clusters derived from the extracted features, are visualized in \figref{fig:cluster_vs_pca_plot}, providing a clear representation of the clustering structure and its alignment with the underlying data characteristics.

The clustering results revealed meaningful groupings that enable tailored predictive modeling and operational optimization. For example, cluster 0 aggregates \glspl{ap} characterized by high activity levels and significant variability between weekdays and weekends. This cluster presents unique challenges for prediction due to its dynamic nature, highlighting the need for advanced forecasting techniques to address such variability. Conversely, clusters 2 and 4 consist of \glspl{ap} with low activity levels, making them ideal candidates for energy-saving strategies. These clusters could benefit from operations such as turning off \glspl{ap} during inactive nighttime periods, demonstrating the practical implications of clustering for resource optimization in managed Wi-Fi networks.

In addition to enabling tailored predictive modeling, the clustering technique shows promise for anomaly detection. By analyzing cluster density, unusual time series can be identified based on their low average similarity scores or their inclusion in the smallest cluster. This capability provides a foundation for detecting anomalous behaviors within the network, which could be further explored in future work. The ability to leverage clustering for both predictive modeling and anomaly detection underscores the versatility and potential impact of the proposed framework.

These results highlight the practical significance of feature-based clustering in addressing challenges related to prediction, energy efficiency, and anomaly detection in large-scale managed Wi-Fi networks. The insights gained from clustering pave the way for more efficient and adaptive network management strategies, offering a scalable solution to the complexities of operational data analysis.

% \subsection{Global Model Creation}

% initial results are obtained using the global model $\gls{lstm}_{gl}$, trained across all time series in the dataset.
% Also, an individual model ($\gls{lstm}_{cl}$) prediction model is produced training the same \gls{lstm} model for each cluster obtained.
% The baseline LSTM algorithm, where no subgrouping is performed, but a model is generated globally across all time series in the dataset.
% We consider a baseline \gls{lstm} model, a type of \gls{rnn} designed to address long-range dependencies in sequential data thanks to the usage of \gls{lstm} cells, which allow capturing temporal dependencies across time series data.
% In this work, the time series of network data are passed through a few \gls{lstm} layers before a predictive output is generated by a feed-forward fully-connected layer~\cite{wilhelmi2024ai}.

\subsection{Cluster-models evaluation}
\begin{figure}[!t]
  \centering
  \subfloat[10 min]{
    \includegraphics[width=.95\linewidth]{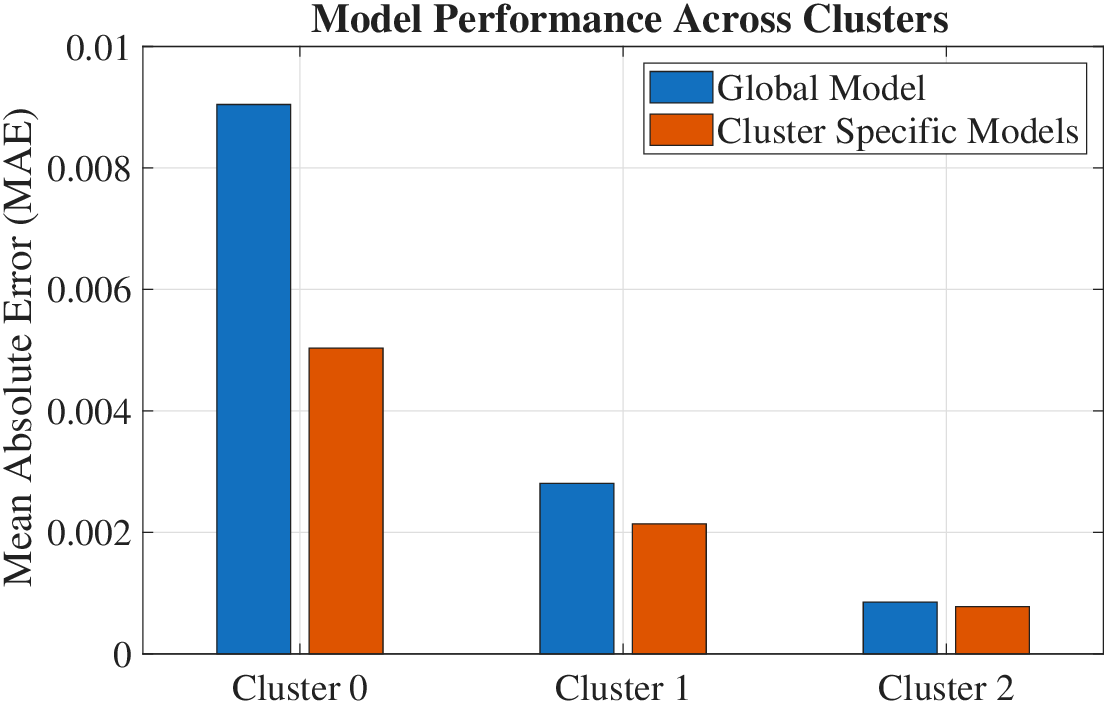}
  }
  
  \subfloat[60 min]{
    \includegraphics[width=.95\columnwidth]{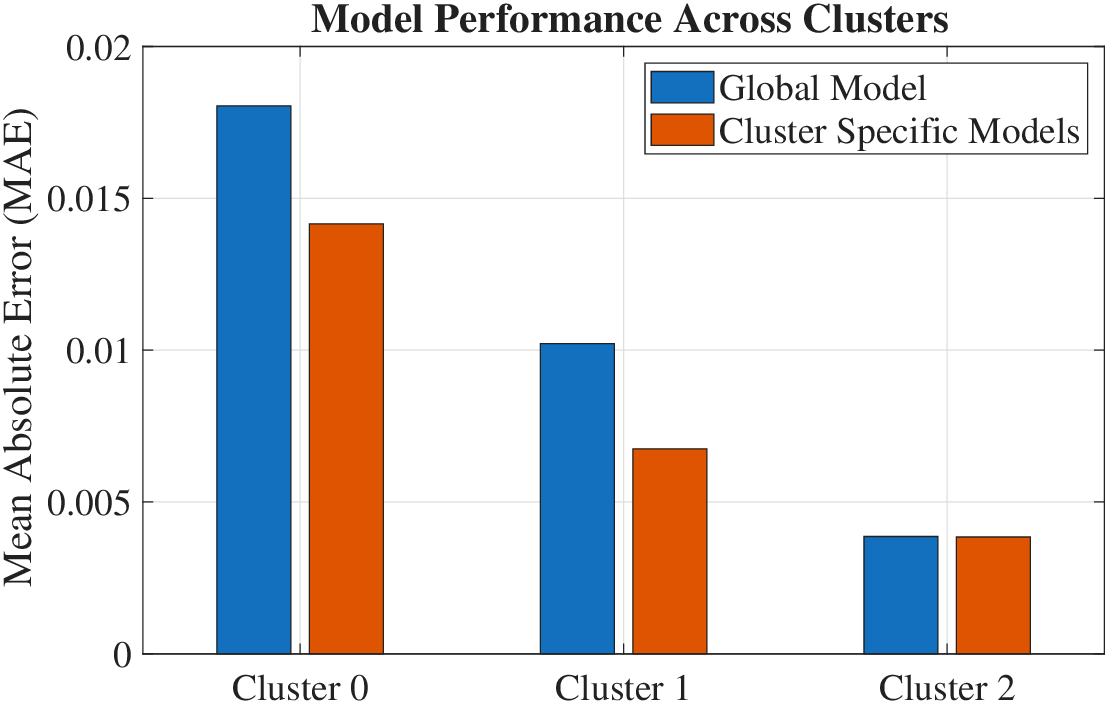}
  }
  \caption{10 min (a) and 60 min (b) future prediction for different clusters of data in a Wi-Fi campus network. The figure shows the results for a global and specialized model.}
  \label{fig:specialized_vs_globalMAE}
\end{figure}

% Once the clusters are obtained,  each cluster. In this section, we provide a description and preliminary performance assessment of the method described in \ref{Procedure}. Following Algorithm \ref{alg:Procedure}, GM inference over the $K = 5$ clusters was performed, using a global \gls{lstm} model ($\gls{lstm}_{gm}$.) based on prior findings \cite{wilhelmi2024ai}.
The GM model, composed of three layers with fifty neurons per layer, was designed for inference across K=5 clusters and evaluated using \gls{mae}. The results, summarized in Table \ref{tab:PerformanceTable}, provide a comprehensive overview of the model's performance across all clusters. While GM demonstrated adequate performance for certain clusters, its accuracy varied significantly across others, motivating the need for cluster-specific models to address these discrepancies.

To enhance prediction accuracy within individual clusters, cluster-specific models ($L_k$) were trained for future 10 and 60-minute predictions.
Figure \ref{fig:specialized_vs_globalMAE} compares the \gls{mae} performance of GM and $L_k$, demonstrating the advantages of cluster-specific training for certain clusters while retaining the resource efficiency of the global model for others.
% based on predefined thresholds, such as $\gls{mae}_{th}$ for future 10-minute predictions. 
Note that, although the MAE values appear to be very low (below 1$\%$), when we multiply it by the size of the network, it results in a relevant number of affected units.
Analyzing the first three clusters, while GM performed sufficiently for cluster 2, the cluster-specific model $L_k$ significantly improved accuracy for cluster 1. 

\begin{figure}[t!]
\centering
\includegraphics[width=0.99\linewidth]{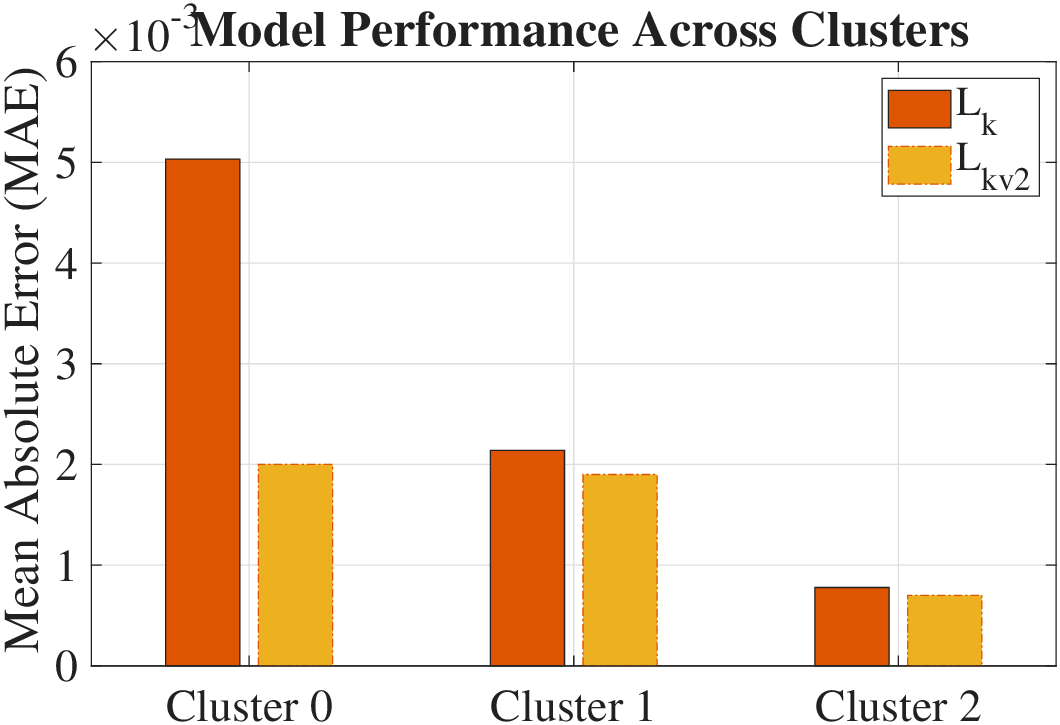}
\vspace{0.1cm}
    \caption{Improvement on MAE using $L_{kv2}$ (LSTMv2) model vs $L_{k}$ (LSTM) (10 min case only).}
    \label{fig:specializedv2}
\end{figure}

Cluster 0 posed unique challenges due to its high activity levels, numerous connected users, high traffic, and significant variability, as highlighted in \figref{fig:cluster_vs_pca_plot}. Both the GM and cluster-specific model $L_k$ resulted in higher \gls{mae} values for this cluster, indicating the need for a more sophisticated approach. To address this, a more complex model, $L_{kv2}$, was trained on cluster 0. This model incorporates two additional layers and increases the number of neurons per layer to 200. As illustrated in \figref{fig:specializedv2}, $L_{kv2}$ achieved a substantial improvement in \gls{mae}, reducing prediction error by 60$\%$ for cluster 0. In contrast, clusters 1 and 2 exhibited only a 10$\%$ improvement when $L_{kv2}$ was applied.
Thus, although models that are more complex and resource-demanding provide greater capacity to capture the intricate dynamics of specific clusters, they should be deployed exclusively for clusters requiring such resources.

% figure lstmv2 for cluster0
%improvement of 60% for cluster 0, 10% for other clusters
% \figref{fig:specializedv2} illustrates the improvement in \gls{mae} achieved by selecting $\gls{lstm}_{clv2}$ as the trained model for the specific cluster $0$. 
% Although clusters 1 and 2 were not selected, their inference results are also presented for comparison. The results show that cluster 0 achieves a significant performance improvement of 60$\%$, whereas the other clusters exhibit only a 10$\%$ improvement when $\gls{lstm}_{clv2}$ is trained on those clusters. This highlights the rationale for employing a more complex model, which demands greater storage and computational resources, exclusively for clusters that require such resources to achieve substantial performance gains.

% show accuracy improvement
\begin{table}[h]
    \centering
    \small % Reduce font size
    \setlength{\tabcolsep}{3.5pt} % Adjust column spacing
    \begin{tabular}{lccccc}
    \hline
    KPI & C0 & C1 & C2 & C3 & C4 \\
    \hline
    MAE GM(10 min) & 0.009 & 0.0028 & 0.00085 & 0.00327 & 0.00064 \\
    MAE GM(60 min) & 0.018 & 0.010 & 0.0039 & 0.0073 & 0.00043 \\
    MAE (10 min) & 0.0050 & 0.0021 & 0.0008 & 0.0013 & 0.0003 \\
    MAE (60 min) & 0.014 & 0.006 & 0.004 & 0.0044 & 0.00044 \\
    % False Positive (\%) & 0.2 & 0.33 & 0.23 & 0.28 & 0.18 \\
    \small{99\%tile (10 min)} & 0.05 & 0.036 & 0.005 & 0.018 & 0.01 \\
     \small{99\%tile (60 min)} & 0.22 & 0.16  & 0.028 & 0.1 & 0.02 \\
    \hline
    \end{tabular}
    \caption{KPI results for GM and $L_k$ across all clusters. The error in the $99\%$tile is in MB. }
    \label{tab:PerformanceTable}
\end{table}

Finally, by selectively applying $L_{kv2}$ only to cluster 0, the framework demonstrates scalable resource allocation: computationally intensive models are deployed only where their complexity is justified by significant accuracy gains, 
while simpler clusters rely on the resource-efficient GM. Considering only the model size of the models, the GM requires (approximately) only 1 MB of memory of storage (deployed once in the central controller).
$L_k$ demands also 1 MB, that in case of using only tailored models, should be multiplied for the number of clusters. 
The most memory hungry model, $L_{kv2}$ and 3.5 MB, is limited to cluster 0. 
In Table \ref{tab:ComputationalSaving} we have reported the storage needed and the accuracy (averaged over the five clusters) for three cases: 1) only one GM deployed on the controller, 2) only cluster specific models (all $L_{kv2}$), 3) an energy oriented approach that finds a  trade-offs between model accuracy and memory resource utilization (one GM, $L_{kv2}$ for cluster 0, $L_k$ for cluster 1 and 3).
Note that cluster-specific models may have higher memory footprints than 
the GM (due to specialized architecture), but their deployment is restricted to clusters exhibiting complex dynamics that justify the additional resource investment.
The system reduces overall memory usage by approximately $40\%$ compared to using only cluster specific models while performing similarly to the best case, ensuring efficient resource allocation without compromising prediction accuracy. This selective approach demonstrates the scalability and practicality of the framework, especially in large-scale networks where resource constraints are critical.

\begin{table}
    \centering
    \begin{tabular}{c|ccc}
    \hline
      Models   & Deployed   & Storage  & Average\\
      Deployed  &  Models &   &  Accuracy\\
      \hline
       One Global Model  & 1 & 1 MB & 0.008\\%&\\
       All Cluster-Specific ($L_{kv2}$)   & 5 & 17.5 MB & 0.004 \\%&\\
       Scalable Deployment Strategy & 3 & 5.5 MB & 0.0044\\%& \\
       \hline
    \end{tabular}
    \caption{Summary of the models cost in term of memory and computations.}
    \label{tab:ComputationalSaving}
\end{table}

% These results highlight the importance of balancing model complexity with resource efficiency, providing a scalable solution for predictive modeling in managed Wi-Fi networks. 
% By tailoring model deployment to cluster-specific needs, the framework achieves significant performance improvements while maintaining efficient resource utilization, paving the way for adaptive and cost-effective network management strategies.

\section{Conclusion}
This manuscript establishes the critical role of clustering and tailored predictive modeling in optimizing managed Wi-Fi networks. 
The integration of feature-based clustering algorithms, supported by \gls{pca} and feature engineering, enables efficient grouping of time series data, facilitating the deployment of resource-optimized predictive models.
The comparative analysis demonstrates that, in high-activity clusters, cluster-specific models outperform global models in terms of prediction accuracy, with significantly lower \gls{mae} values observed. The study highlights the trade-offs between model complexity (and accuracy) and 
computational resource utilization, emphasizing the need for scalable solutions that balance predictive accuracy with adaptive resource allocation across network clusters. In future work, we will investigate possible developments in PCA, such as the use of LDA and embedding space definitions, to counteract the effects of missing data.
% Adaptive network management strategies, which prioritize cluster-specific requirements, are identified as essential for optimizing resource allocation and ensuring reliability in dynamic network environments. These findings provide a robust framework for enhancing predictive modeling and resource efficiency, paving the way for sustainable and scalable advancements in managed Wi-Fi network operations.

%%%%%%%%%%%%%%%%%%%%%%%%%
%%%  BIBLIOGRAPHY    
%%%%%%%%%%%%%%%%%%%%%%%%%
%\bibliographystyle{unsrt}
\bibliographystyle{IEEEtran}
\bibliography{biblio}

\end{document}